# Angle-insensitive spectral imaging based on topology-optimized plasmonic metasurfaces

**Jiawei Yang,**[1,2] **Kaiyu Cui,**[1,2,*] **Yidong Huang,**[1,2,3,*] **Wei Zhang,**[1,2,3] **Xue Feng,**[1,2] **and Fang Liu**[1,2]

[1]*Department of Electronic Engineering, Tsinghua University, Beijing 100084, China*
[2]*Beijing National Research Center for Information Science and Technology (BNRist), Tsinghua University, Beijing 100084, China*
[3]*Bejing Academy of Quantum Information Science, Beijing 100084, China*
*\*Corresponding author: kaiyucui@tsinghua.edu.cn, yidonghuang@tsinghua.edu.cn*

**Abstract:** On-chip spectral imaging based on engineered spectral modulation and computational spectral reconstruction provides a promising scheme for portable spectral cameras. However, the angle dependence of modulation units results in the angle sensitivity of spectral imaging, which limits its practical applications. Here, we proposed a design for angle-robust spectral recovery based on a group of topology-optimized plasmonic metasurface units under a 30 ° field-of-view, and demonstrate angle-insensitive on-chip spectral imaging in the wavelength range of 450 to 750 nm for average polarization. Furthermore, we experimentally verified the angle-insensitive spectral filtering effects of the fabricated metasurface units, and demonstrated angle-robust spectral reconstruction with a fidelity of 98%. Our approach expands the application scale of spectral imaging, and provides a guidance for designing other angle-robust devices.

## 1. INTRODUCTION

Spectral imaging acquires images across a wide range of electromagnetic spectrum, finding its applications in numerous fields such as precision agriculture [1], disease diagnosis [2, 3], food quality control [4], environmental monitoring [5], and art conservation [6]. The majority of spectral imaging devices are based on spatial or spectral scanning, which requires long acquisition time and may produce undesirable artifacts for motional objects. While for snapshot spectral imagers, real-time dynamic spectral image information can be captured without moving components [7]. Besides, the increasing demand for portable or handheld spectral cameras pushes the integration and miniaturization of spectral imaging devices. Recently, computational spectral reconstruction techniques based on engineered spectral response characteristics become a research hotspot. Related schemes include quantum dots [8, 9], disordered scattering structure [10], nanowire [11, 12], photonic crystal slabs [13, 14] and so on. For on-chip snapshot spectral imaging, successful prototypes have been reported, including a 10×10 pixel hyperspectral imager based on photonic-crystal slabs [13] and a 72×88 pixel ultraspectral imager based on silicon metasurfaces [15, 16]. However, due to the angular dispersion of the spectral modulation units, the incident angle is required to be the same for the calibration and measurement, which hinders the practical applications of on-chip spectral imaging. Besides, there also exist commercial spectral cameras (e.g., Imec [17] and Pixelteq [18]) based on on-chip thin-film Fabry-Perot filter arrays, which are also angle sensitive due to that the path lengths of interfering beams vary with the angle of incidence. Although there are methods to predict and correct for the angular dependency for camera lenses without vignetting [19-21] or with vignetting [22], these models are developed only for narrowband thin-film Fabry-Perot filters, and only spectral shift is corrected with the assumption of invariant filter bandwidth in these works [19-22].

The physical origin for angle-dependent spectral response is the momentum matching condition. To address this issue, several approaches have been proposed and applied to realize angle robust color filtering. In 2013, Yi-Kuei Ryan Wu proposed to utilize the metal-insulator-metal Fabry-Perot (MIMFP) cavity modes in metallic nanoslits to achieve angle-insensitive structural colors [23]. In 2016, Chenying Yang proposed a new scheme using a single layer of ultrathin metal patch array structure, where the angle insensitive color filtering feature is attributed to the localized surface plasmon resonance (LSPR) excited within the structure [24]. Besides, thin film structures are investigated as an alternative method, such as the Fabry-Perot cavity structure with high refractive index material as the spacing dielectric layer [25, 26], ultrathin highly absorptive dielectric film on a metallic substrate [27], and a dielectric layer sandwiched by two metal films on a fused silica substrate [28-31]. However, the above researches are confined to structural colors due to limited degrees of freedom in structure design. As for spectral reconstruction, a large number of angle-insensitive distinctive spectral filtering functions with as low mutual correlation as possible are required [13], which has not been realized and reported yet.

In this work, we propose to employ topology-optimized plasmonic metasurface units to generate diverse angle-insensitive spectral response functions for computational spectral recovery. In order to expand the metasurface design space for diverse spectral responses, we get rid of the traditional methods of metasurface design based on regular shapes and develop an inverse design approach based on freeform shaped meta-atoms. By establishing a large geometric library containing thousands of different shapes with proper feature sizes and adopting genetic algorithm, both the period and shape of the meta-atom (i.e., the unit cell of each metasurface unit) are optimized. Different from other inverse design methods such as adjoint topology optimization [32-35] or deep neural network [36-39], our method combines a global optimizer and a freeform shape generator which takes the fabrication constraints into account. The meta-atoms can be classified into the hole-type and pillar-type according to their shapes. Through the simulation of electric field intensity distributions, we show that the angle-insensitive features are attributed to the MIMFP cavity modes in the hole-type structures as well as the LSPR modes in the pillar-type structures. As a proof of principle, we designed 100 plasmonic metasurface units using gold material for angle robust spectral reconstruction, and further demonstrate angle-insensitive spectral imaging based on simulated spectral reconstruction using an auto-encoder. Experimentally, we calibrated the transmission spectra of the fabricated metasurface units, and measured the spectrum of a LED light signal, showing angle-insensitive spectral reconstruction. The proposed scheme has great potential for various applications of spectral imaging, and provides a promising method for designing a wider variety of angle-robust devices.

## 2. RESULTS

The schematic geometry of the proposed structure integrated with a CMOS imaging sensor (CIS) is depicted in Fig. 1(a), and the vertical section is shown in Fig. 1(b). An ultrathin gold metasurface layer of 20 nm is built on a quartz substrate and covered by a layer of spin on glass (SOG). The layer of SOG is used to avoid shorting with metallic pads on the CIS [40]. The plasmonic layer consists of multiple periodic array units with different periods and shapes, producing distinctive spectral response characteristics. Therefore, the photodetector underneath each metasurface unit receives different signal, from which the spectrum of incident light can be reconstructed. As shown in Fig. 1(c), suppose $N$ metasurface units are used for spectrum recovery, and the unknown spectrum $f(\lambda)$ is discretized into an $M$-dimension vector, then the detected signals can be summarized as:

$$I = Tf + e \qquad (1)$$

where $I$ is the $N \times 1$ measurement vector, $T$ is the $N \times M$ transmission spectrum matrix, $f$ is the $M \times 1$ unknown spectrum vector, and $e$ is the $N \times 1$ measurement noise vector. The target spectrum $f(\lambda)$ can be recovered by solving Eq. (1). Henceforth, the $N$ metasurface units form a micro-spectrometer, and an array of identical micro-spectrometers can be used for spectral imaging.

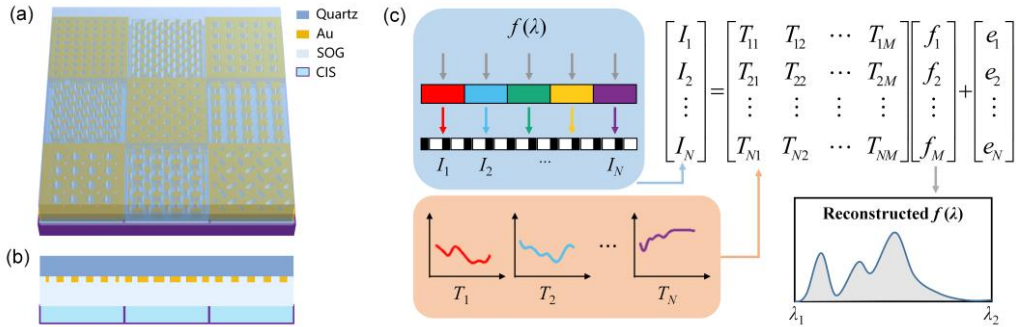

**Fig. 1** (a) A schematic view of the proposed structure integrated with a CMOS image sensor (CIS). The structure consists of a 20-nm thick gold metasurface layer on a quartz substrate and a cover layer of spin on glass (SOG). (b) The vertical section of the structure in (a). (c) Principles of the computational spectral reconstruction based on the proposed structure.

For the purpose of spectral reconstruction, we need to obtain diverse spectral modulation functions by tuning the structural parameters. To break through the limitation of regular shapes in traditional metasurface design [13-15, 24] and further expand the design space, we developed an algorithm for generating a great number of freeform shapes with controllable feature sizes (see Supplementary Material S1 for details), and optimize both the period and shape of each meta-atom.

To explore the incident angle dependence of the metasurface units with freeform shaped meta-atoms, we simulated the angle resolved transmission spectra of a gold hole array as shown in Fig. 2(a) and a gold pillar array in Fig. 2(e) for average polarization using Reticolo V9 [41], an open source software based on rigorous coupled-wave analysis (RCWA) [42]. In the simulation, the complex refractive index of gold is taken from the data in Ref. [43], while quartz and SOG have a constant refractive index of 1.52 and 1.45, respectively. From the results in Fig. 2(b)(f), it can be seen that the transmission spectrum almost remains the same when the incident angle varies from 0 ° to 30 °. Furthermore, we simulated the electric field profile of the corresponding structure for the wavelength of 675 nm and 687 nm as shown in Fig. 2(c)(g), respectively. For the gold hole array structure, the electric field is well confined within the hole region corresponding to a MIMFP cavity mode which is an angle independent mode [23]. Fig. 2(g) shows the gold pillar array structure supports LSPR modes [24], which is excited by appropriate polarization and frequency of the incident light regardless of the incident angle.

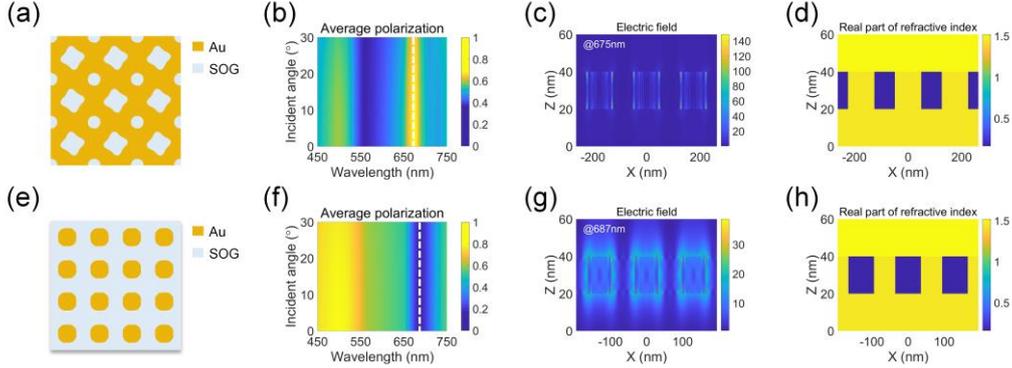

**Fig. 2** (a) A gold hole array structure with the period of 175 nm, and (b) its transmission spectra for average polarization with the incident angle varying from 0 ° to 30 °. (c) The corresponding simulated electric field distribution for the wavelength of 675 nm (indicated by the white dashed line in (b)) as well as (d) the real part of the refractive index distribution. (e) A gold pillar array structure with the period of 130 nm, and (f) its transmission spectra for average polarization with the incident angle varying from 0 ° to 30 °. (g) The corresponding simulated electric field distribution for the wavelength of 687 nm (indicated by the white dashed line in (f)) as well as (h) the real part of the refractive index distribution.

To obtain a group of angle-robust transmission spectra for spectral reconstruction, we set the design objective as:

$$\text{minimize} \left( \frac{1}{N} \sum_{i=1}^{N} \frac{\|T_{i1} - T_{i2}\|_1}{\|T_{i1}\|_1} - \alpha \cdot \frac{1}{N} \sum_{i=1}^{N} \min_{i \neq j} \|\tilde{T}_{i1} - \tilde{T}_{j1}\|_2 \right), \quad \alpha > 0 \tag{2}$$

where $N$ is the number of transmission spectra, $\alpha$ is the weight factor, $T_{i1}$, $T_{i2}$ are the transmission spectra of the $i$th metasurface unit at the incident angle of 0 ° and 15 °, respectively, and $\tilde{T}_{i1}, \tilde{T}_{j1}$ are the $l_2$-normalized transmission spectra of the $i$th, $j$th metasurface units under normal incidence, respectively. Here, the average relative difference of the transmission spectra at the incident angle of 0 ° and 15 ° is defined by the first part in Eq. (2), which represents the angle sensitivity. Here, we choose an angle of 15 ° for a proof-of-principle demonstration, corresponding to a 30 ° field-of-view, which is also a typical value for most camera lenses. The second part in Eq. (2) means the mean minimum mutual differences among the $N$ spectra, which is maximized for improving the performance of spectral reconstruction.

The optimization process is illustrated in Fig. 3(a). With the aid of the shape-generation algorithm, we first build a geometric library containing thousands of freeform shapes, and pick those with feature size larger than 50 nm with the period varying from 120 to 300 nm in a step of 5 nm. Then, the transmission spectra for all the selected shapes under each period are simulated via RCWA. After that, we apply genetic algorithm to find the optimal $N$ pairs of period and shape number, where a population of individuals is first initialized by random sampling, and the fitness scores of corresponding transmission spectra are calculated, followed by an iteration process of selection, reproduction, mutation and score evaluation until the fitness scores converge.

After optimization, we obtain 100 angle-insensitive transmission spectra as shown in Fig. 3(b). For comparison, we randomly select 100 220-nm c-Si metasurface units with freeform-shaped meta-atoms, and simulate their transmission spectra as shown in Fig. 3(c). As manifested in Fig. 3(d), the average relative difference at the incident angle of 0 ° and 15 ° (i.e., angle sensitivity) is merely 0.98% for the optimized angle-insensitive transmission spectra, while the transmission spectra for the c-Si metasurfaces exhibit an angle sensitivity as high as 23.18%. Furthermore, for the same average relative difference of 10%, the angle tolerance is up to 48 ° for the angle-insensitive transmission spectra, much larger than that for the angle-sensitive transmission spectra, where the angle tolerance is only 6 ° as indicated by the dashed line in Fig. 3(e). In addition, to show the generality of our method, we also designed 100 metasurface units based on silver material (see Supplementary Material S3 for details).

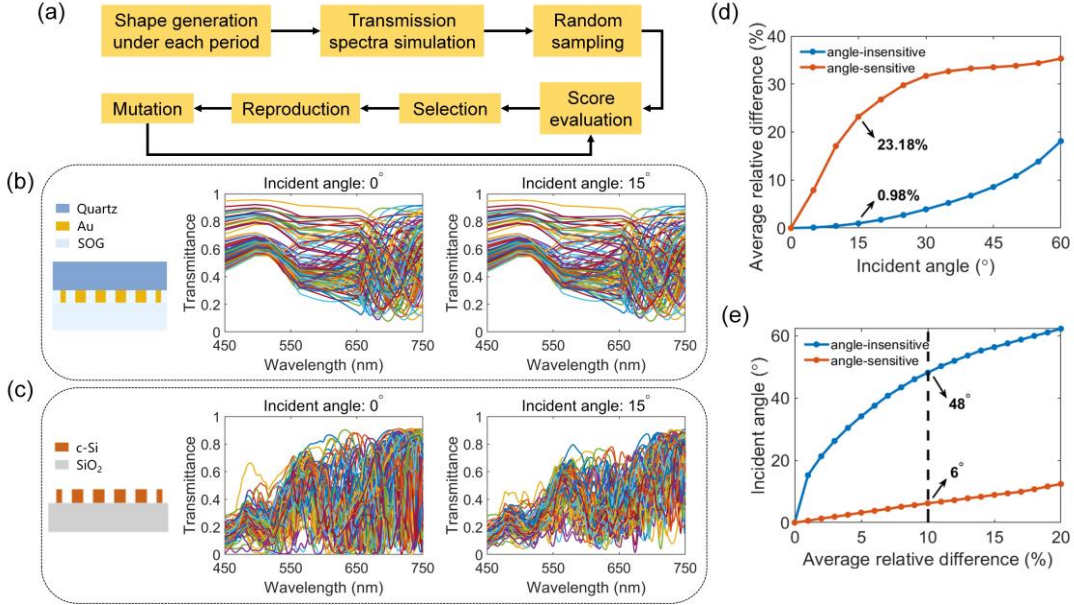

**Fig. 3** (a) The flowchart of the optimization process with a loop of score evaluation, selection, reproduction and mutation. (b) The optimized 100 angle-insensitive transmission spectra for the proposed structure at the incident angle of 0 ° and 15 °. (c) Simulated transmission spectra of 100 silicon-based metasurface units at the incident angle of 0 ° and 15 °. (d) The average relative difference of the angle-insensitive and angle-sensitive transmission spectra as a function of incident angle compared with the case for normal incidence. (e) The incident angle for different average relative difference of the angle-insensitive and angle-sensitive transmission spectra.

Furthermore, to test the performance of spectral reconstruction using the optimized transmission spectra, we implement simulated spectral reconstruction through an auto-encoder as illustrated in Fig. 4(a), which is composed of an encoder and a decoder. Here, the encoder contains only an unbiased fully connected layer with the weight matrix corresponding exactly to the transmission spectra [44]. Therefore, the encoder plays the role of simulated measurement with the input spectrum $f$ and output measurement vector $I$. Besides, we add Gaussian white noise [45, 46] to $I$ as measurement noise vector $e$ of Eq. (1). As for the decoder, it consists of three fully connected layers for spectral recovery. The number of input and output nodes for each layer is marked in Fig. 4(a). We build a synthetic spectral dataset for training the network, where the number of training and test sets are randomly split by a ratio of 9:1. The dataset contains 200,000 Gaussian line shape spectra, which are produced by adding a series of Gaussian distribution component functions together [45]. The network is trained for 100 epochs using Adam optimizer with the batch size of 2000. The initial learning rate is 0.001 and decay 1.25 times every 10 epochs.

To evaluate the performance of spectral reconstruction, we use a metric of fidelity defined as:

$$F(f_1, f_2) = \langle f_1, f_2 \rangle \tag{3}$$

where $f_1$, $f_2$ are the $l_2$-normalized original spectrum and reconstructed spectrum vector, respectively, and $\langle \rangle$ means the inner product. Under different noise levels, we train the network using the transmission spectra under normal incidence in Fig. 3(b) and Fig. 3(c). Then, we calculate the average fidelities of spectral reconstruction on the test set under the incident angle of 0 ° and 15 °, using the trained network. For the optimized angle-insensitive transmission spectra, it can be seen that there is no significant reduction in the fidelities when the incident angle varies from 0 ° to 15 ° in Fig. 4(b), whereas the case is different for the angle sensitive transmission spectra of c-Si metasurface units in Fig. 4(c).

For the demonstration of spectral imaging, we use the trained spectral auto-encoder to reconstruct the data cube of 254×345×301 for a 24-patch Macbeth color checker with 2% Gaussian white noise, where the 2% level is just an example. Fig. 4(d) shows the post-colored image in RGB form obtained by a commercial spectral camera (Dualix Instruments, GaiaField Pro V10) as a reference. The reconstructed results using the optimized simulated transmission spectra are displayed in Fig. 4(e) and (f), where the reconstructed image hardly varies with the incident angle and shows angle-insensitive performance. As for the results in Fig. 4(h) and (i) using the transmission spectra in Fig. 3(c), the performance of spectral imaging significantly deteriorates as the incident angle increases. Through the sampled spectra in Fig. 4(g) and (j) for the point *P* in Fig. 4(d), it is confirmed again that the optimized transmission spectra can be used for angle-robust spectral imaging. The results of simulated spectral imaging using the designed silver metasurface units are provided in the Supplementary Material S3.

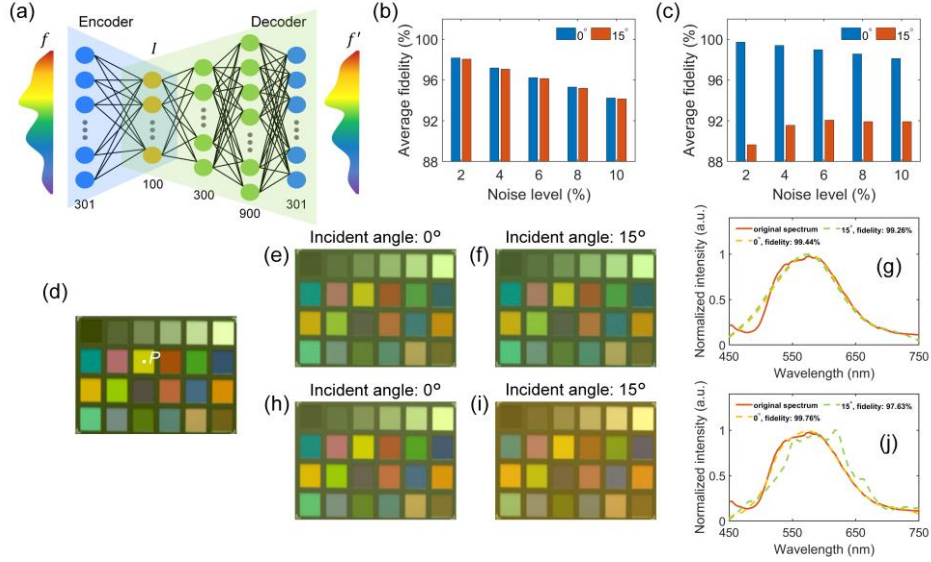

**Fig. 4** (a) The network architecture of the auto-encoder for simulated spectral reconstruction. (b) and (c) Average spectral reconstruction fidelity for the test spectral dataset using the transmission spectra in Fig. 3(b) and Fig. 3(c) as the weight matrix of the encoder in (a), respectively. (d) The post-colored image for a 24-patch Macbeth color chart obtained using a commercial spectral camera (Dualix Instruments, GaiaField Pro V10). (e, f) and (h, i) The reconstructed post-colored images with 2% Gaussian white noise using the transmission spectra in Fig. 3(b) and Fig. 3(c), respectively. (e)(h) correspond to the incident angle of 0 ° and (f)(i) correspond to 15 °. (g) and (j) Results of simulated spectral reconstruction with 2% Gaussian white noise for the point *P* in (d) using the transmission spectra in Fig. 3(b) and Fig. 3(c), respectively.

To experimentally validate the proposed approach, we fabricated 900 gold metasurface units (see Supplementary Material S2 for details), as indicated in Fig. 5(a). The scanning electron microscope (SEM) images of four selected metasurface units are presented in Fig. 5(b). We calibrated the transmission spectra at different incident angles in the wavelength range of 450 to 750 nm. The angle-insensitive feature can be clearly seen from the results presented in Fig. 5(c) for the four metasurface units in Fig. 5(b). Quantitatively speaking, the average relative difference of the 100 calibrated transmssion spectra at the incident angle of 0 ° and 15 ° is only 3.72%. To show the angle-robust spectral reconstruction, we measured the spectrum of a LED light source at the incident angle of 0 ° and 15 °, with the reference spectrum obtained by a commercial spectrometer (OceanView QE Pro). From the results in Fig. 5(d), we can see that the spectrum can be roughly recovered with a fidelity around 98%, even under the incident angle of 15 °. The performance for spectral reconstruction can be further improved by increasing the mutual differences among the 100 transmission spectra, which can be achieved by using multilayer structure, decreasing the feature size, using other metal materials, etc. Besides, in principle, angle-insensitive snapshot spectral imaging can be realized by fabricating a larger number of metasurface units.

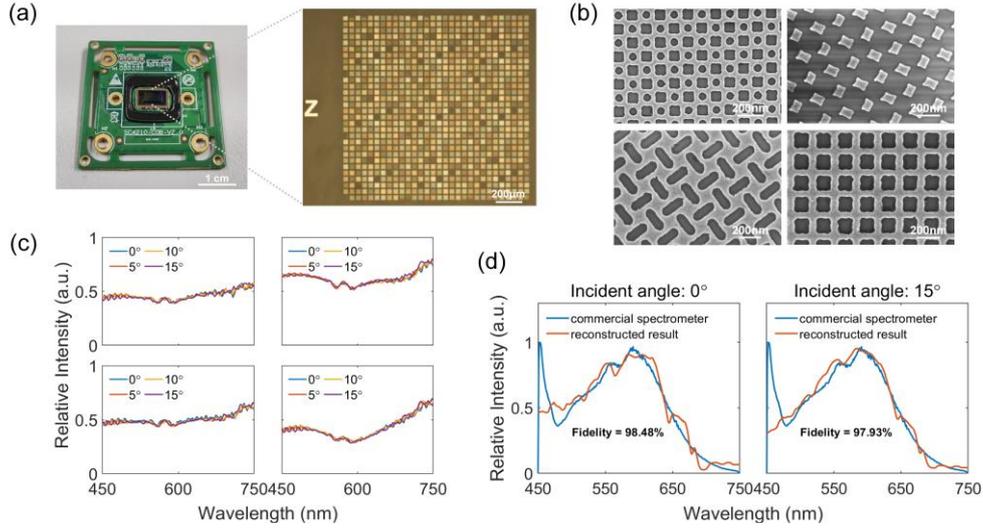

**Fig. 5** (a) Device photograph and optical image of the fabricated metasurface units. (b) Scanning electron microscope (SEM) images of four metasurface units. (c) The calibrated transmission spectra of the four metasurface units in (b) at the incident angle of 0°, 5°, 10° and 15°. (d) Measurements of the spectrum of a LED at the incident angle of 0° and 15°.

## 3. CONCLUSIONS

In conclusion, we propose to utilize the topology-optimized plasmonic metasurfaces as spectral modulation units to address the sensitivity of transmission spectra to the incident angle. Through the joint optimization of the period and shape for each meta-atom, we designed 100 plasmonic metasurface units with distinctive angle-robust transmission spectra whose angle sensitivity is merely 0.98%, and the angle tolerance for the average relative difference of 10% is up to 48°, which is significantly improved compared to the silicon metasurfaces. Furthermore, we test the performance of simulated spectral reconstruction by training an auto-encoder with a dataset of 200,000 synthetic Gaussian line shape spectra, and demonstrate angle-insensitive spectral imaging for a standard color calibration board. Finally, we fabricated a group of metasurface units, and experimentally realized angle insensitive spectral reconstruction. We believe that our approach has great potential to push the development, and expand the application fields of spectral imaging, such as large field-of-view or off-axis spectral imaging. Finally, beyond the field of spectral imaging, the methods proposed here can also be used as a guideline for designing other angle-robust devices.

**Data and materials availability:** The data and materials that support the findings of this study and custom codes are available from the corresponding author upon reasonable request.

**Acknowledgments:** The authors would like to thank Tianjin H-Chip Technology Group Corporation, Innovation Center of Advanced Optoelectronic Chip and Institute for Electronics and Information Technology in Tianjin, Tsinghua University for their fabrication support with EBL and ICP etching.

**Funding:** The National Key Research and Development Program of China (2018YFB2200402); Beijing Municipal Science and Technology Commission (Z201100004020010); Beijing Frontier Science Center for Quantum Information; and Beijing Academy of Quantum Information Sciences.

**Disclosures:** The authors declare no conflicts of interest.

**Author contributions:** J. Y. conceived the study, performed the simulation and implemented the experiment. K.C. and Y.H. supervised the project and provided much support on device fabrication and characterization. J.Y. and K.C. wrote the manuscript. Y. H., W. Z., X. F. and F. L. discussed the results and reviewed the paper.

# REFERENCES

1. Lebourgeois, V.; Bégué, A.; Labbé, S.; Mallavan, B.; Prévot, L.; Roux, B. Can commercial digital cameras be used as multispectral sensors? A crop monitoring test. *Sensors* **2008**, 8, 7300-7322.
2. Lowe, A.; Harrison, N.; French, A. P. Hyperspectral image analysis techniques for the detection and classification of the early onset of plant disease and stress. *Plant methods* **2017**, 13, 1-12.
3. Lu, G.; Fei, B. Medical hyperspectral imaging: a review. *J Biomed Opt* **2014**, 19, 010901.
4. Feng, Y. Z.; Sun, D. W. Application of hyperspectral imaging in food safety inspection and control: a review. *Crit Rev Food Sci Nutr* **2012**, 52, 1039–1058.
5. Stuart, M. B.; McGonigle, A. J.; Willmott, J. R. Hyperspectral Imaging in Environmental Monitoring: A Review of Recent Developments and Technological Advances in Compact Field Deployable Systems. *Sensors* **2019**, 19, 3071.
6. Liang, H. Advances in multispectral and hyperspectral imaging for archaeology and art conservation. *Appl. Phys. A* **2012**, 106, 309–323.
7. Hagen, N. A.; Kudenov, M. W. Review of snapshot spectral imaging technologies. *Opt. Eng.* **2013**, 52, 090901.
8. Bao, J.; Bawendi, M. G. A colloidal quantum dot spectrometer. *Nature* **2015**, 523, 67-70.
9. Zhu, X.; et al. Broadband perovskite quantum dot spectrometer beyond human visual resolution. *Light Sci. Appl.* **2020**, 9, 1-9.
10. Redding, B.; Liew, S. F.; Sarma, R.; Cao, H. Compact spectrometer based on a disordered photonic chip. *Nat. Photonics* **2013**, 7, 746-751.
11. Yang, Z.; et al. Single-nanowire spectrometers. *Science* **2019**, 365, 1017–1020.
12. Meng, J.; Cadusch, J. J.; Crozier, K. B. Detector-only spectrometer based on structurally colored silicon nanowires and a reconstruction algorithm. *Nano Lett.* **2019**, 20, 320–328.
13. Wang, Z.; et al. Single-shot on-chip spectral sensors based on photonic crystal slabs. *Nat. Commun.* **2019**, 10, 1-6.
14. Zhu, Y.; Lei, X.; Wang, K. X.; Yu, Z. Compact CMOS spectral sensor for the visible spectrum. *Photonics Res.* **2019**, 7, 961-966.
15. Xiong, J.; et al. Dynamic brain spectrum acquired by a real-time ultraspectral imaging chip with reconfigurable metasurfaces. *Optica*, **2022**, 9, 461-468.
16. Yang, J.; et al. Ultraspectral Imaging Based on Metasurfaces with Freeform Shaped Meta-Atoms. *Laser Photonics Rev.* **2022**, 2100663.
17. Hyperspectral imaging technology imec. https://www.imechyperspectral.com/en/hyperspectral-imaging-technology (accessed 2022-11-25).
18. Spectral Systems Ocean Insight. http://www.pixelteq.com (accessed 2022-11-25).
19. Goossens, T.; Geelen, B.; Pichette, J.; Lambrechts, A.; Van Hoof, C. Finite aperture correction for spectral cameras with integrated thin-film Fabry–Perot filters. *Appl. Opt.* **2018**, 57, 7539-7549.
20. Goossens, T., & Van Hoof, C. Thin-film interference filters illuminated by tilted apertures. *Appl. Opt.* **2020**, 59, A112-A122.
21. Goossens, T.; Vunckx, K.; Lambrechts, A.; Van Hoof, C. Spectral Shift Correction for Fabry-Perot Based Spectral Cameras. In *2019 10th Workshop on Hyperspectral Imaging and Signal Processing: Evolution in Remote Sensing (WHISPERS)* **2019**, 1-6.
22. Goossens, T.; Geelen, B.; Lambrechts, A.; Van Hoof, C. Vignetted-aperture correction for spectral cameras with integrated thin-film Fabry–Perot filters. *Appl. Opt.* **2019**, 58, 1789-1799.
23. Wu, Y. K. R.; Hollowell, A. E.; Zhang, C.; Guo, L. J. Angle-insensitive structural colours based on metallic nanocavities and coloured pixels beyond the diffraction limit. *Sci. Rep.* **2013**, 3, 1-6.

24. Yang, C.; et al. Angle robust reflection/transmission plasmonic filters using ultrathin metal patch array. *Adv. Opt. Mater.* **2016**, 4, 1981-1986.
25. Noh, T. H.; Yoon, Y. T.; Lee, S. S.; Choi, D. Y.; Lim, S. C. Highly angle-tolerant spectral filter based on an etalon resonator incorporating a high index cavity. *J Opt Soc Korea* **2012**, 16, 299-304.
26. Guo, L. J.; Xu, T. Spectrum filtering for visual displays and imaging having minimal angle dependence. US 9261753, 2016.
27. Kats, M. A.; Blanchard, R.; Genevet, P.; Capasso, F. Nanometre optical coatings based on strong interference effects in highly absorbing media. *Nat. Mater.* **2013**, 12, 20-24.
28. Lee, K. T.; Seo, S.; Lee, J. Y.; Guo, L. J. Strong resonance effect in a lossy medium-based optical cavity for angle robust spectrum filters. *Adv. Mater.* **2014**, 26, 6324-6328.
29. Yang, C.; Shen, W.; Zhang, Y.; Li, K.; Fang, X.; Zhang, X.; Liu, X. Compact multilayer film structure for angle insensitive color filtering. *Sci. Rep.* **2015**, 5, 1-5.
30. Mao, K.; Shen, W.; Yang, C.; Fang, X.; Yuan, W.; Zhang, Y.; Liu, X. Angle insensitive color filters in transmission covering the visible region. *Sci. Rep.* **2016**, 6, 1-7.
31. de Souza, I. G.; Rodriguez-Esquerre, V. F.; Rêgo, D. F. Wide-angle filters based on nanoresonators for the visible spectrum. *Appl. Opt.* **2018**, 57, 6755-6759.
32. Sell, D.; Yang, J.; Doshay, S.; Yang, R.; Fan, J. A. Large-angle, multifunctional metagratings based on freeform multimode geometries. *Nano Lett.* **2017**, 17, 3752-3757.
33. Phan, T.; Sell, D.; Wang, E. W.; Doshay, S.; Edee, K.; Yang, J.; Fan, J. A. High-efficiency, large-area, topology-optimized metasurfaces. *Light Sci. Appl.* **2019**, 8, 1-9.
34. Mansouree, M.; McClung, A.; Samudrala, S.; Arbabi, A. Large-scale parametrized metasurface design using adjoint optimization. *ACS Photonics* **2021**, 8, 455-463.
35. Zhou, M.; et al. Inverse design of metasurfaces based on coupled-mode theory and adjoint optimization. *ACS Photonics* **2021**, 8, 2265-2273.
36. Liu, Z.; Zhu, D.; Rodrigues, S. P.; Lee, K. T.; Cai, W. Generative model for the inverse design of metasurfaces. *Nano Lett.* **2018**, 18, 6570-6576.
37. Tanriover, I.; Hadibrata, W.; Scheuer, J.; Aydin, K. Neural networks enabled forward and inverse design of reconfigurable metasurfaces. *Opt. Express* **2021**, 29, 27219-27227.
38. Lininger, A.; Hinczewski, M.; Strangi, G. General Inverse Design of Layered Thin-Film Materials with Convolutional Neural Networks. *ACS Photonics* **2021**, 8, 3641-3650.
39. Lu, X.; et al. Classification and Inverse Design of Metasurface Absorber in Visible Band. *Adv. Theory Simul.* **2022**, 5, 2100338.
40. He, X.; et al. A single sensor based multispectral imaging camera using a narrow spectral band color mosaic integrated on the monochrome CMOS image sensor. *APL Photonics* **2020**, 5, 046104.
41. Hugonin, J. P.; Lalanne, P. Light-in-complex-nanostructures/RETICOLO: V9 (Version 9). Zenodo. https://doi.org/10.5281/zenodo.4419063 (accessed 2022-11-25)
42. Moharam, M. G.; Gaylord, T. K. Rigorous coupled-wave analysis of planar-grating diffraction. *J Opt Soc Am* **1981**, 71, 811-818.
43. Hayness, W. M.; Lide, D. R. *CRC handbook of chemistry and physics*; CRC Press, 2014.
44. Song, H.; et al. Deep-Learned Broadband Encoding Stochastic Filters for Computational Spectroscopic Instruments. *Adv. Theory Simul.* **2021**, 4, 2000299.
45. Kim, C.; Park, D.; Lee, H. N. Compressive sensing spectroscopy using a residual convolutional neural network. *Sensors* **2020**, 20, 594.
46. Lin, X.; Liu, Y.; Wu, J.; Dai, Q. Spatial-spectral encoded compressive hyperspectral imaging. *ACM Trans. Graph.* **2014**, 33, 1-11.